# COULOMB POTENTIAL AND WITTEN'S SUPERALGEBRA

ANZOR KHELASHVILI & TAMAR KHACHIDZE

*/Tbilisi State University, Chavchavadze av. 3. 0128, Tbilisi, Georgia/*

**ABSTRACT:** The additional hidden symmetry of the Coulomb-Kepler problem is reviewed in classical as well as in quantum mechanics. The main purpose is to elucidate the role of this kind of symmetries in the reduction of physical problems, to show algebraic possibilities of derivation of spectra. The original results are presented also. They are hidden symmetries in the Dirac equation, where it is shown that the requirement of invariance of the Dirac Hamiltonian under some kind of Witten's superalgebra, picks out the Coulomb potential only. The problem in the arbitrary higher dimensions is also considered. It is derived that the traditional view on the Coulomb potential is to be changed in the context of N=2 supersymmetry.

## 1. Introduction

Coulomb or Newtonian forces play a decisive role in Nature. They are fundamental in a sense that only they (as well as an isotropic oscillator) give the periodic motion on closed orbits (Bertrand's theorem). The main characteristic feature of these forces is their identical $1/r^2$ dependence on distance or their potential energy behaves as $1/r$.

It is worthwhile to remember that both laws were discovered experimentally. Theoretically the source of their derivation in classical physics is the Gauss theorem, while in quantum theory – the photon propagator, the propagator of massless spin-1 boson, from which the coulomb potential is derived in instantaneous (non-relativistic) approximation.

Moreover, both these approaches in arbitrary dimensional spaces give $r^{2-D}$, where D is the space dimension.

Below we want to suggest the alternate way of derivation of Coulomb potential based on the requirement of N=2 supersymmetry in the Dirac equation [1].

It is well-known from classical mechanics that there is the additional conservation law in the capacity of so-called Laplace-Runge-Lenz (LRL) vector [2-4] besides the angular momentum. Owing to this conservation law the Kepler problem in classical mechanics is solved algebraically without solution of equations of motion. Therefore the degeneracy of orbits occurs which is related to a more higher symmetry than the 3-dimensional rotations. Thanks to this additional conservation law the higher symmetry appears as well in quantum mechanics, which causes, so-called an "accidental" degeneracy of hydrogen atomic levels.

We must make distinction between two kinds of symmetry, namely, geometrical and dynamical.

The geometrical symmetries appear from the invariance to some coordinate transformations. Any transformation fulfills all the axioms of groups: two sequential transformations as an associative group compositions give some transformation anew, there is the identical transformation as a neutral element, and there is the inversed transformation as the inversed element. Therefore it is natural that the geometric symmetries are described by Lie groups and Lie algebras, if these transformations are continuous.

Symmetries in physics are manifested by conservation laws. When we have some symmetry under definite transformation, to each of them there correspond some conserved quantities.

As regards of dynamical symmetries, they, at the first glance, are not related to any transformations and follow directly from equations of motion only for some definite potentials or definite dynamics. They have corresponding conserved quantities, which obey some algebra, but transformations are unknown. Algebras of this origin are named as dynamical
. Their study was very popular in 60-ies of previous century.

## 2. Dynamical Symmetry of the Coulomb Potential in Classical Mechanics

Let us consider the classical Kepler problem of motion of m mass particle in an arbitrary central potential field, when potential energy depends only on distance, $V = V(r)$. As is well known, the angular momentum $\vec{L} = [\vec{r} \times \vec{p}]$, $\vec{p} = m\vec{v}$, is conserved in these fields. It follows from the invariance under 3-space rotations (isotropy of space). It is the geometric symmetry as it is connected to coordinate transformations. This law allows to separate angular variables and to reduce the problem to the single-dimensional one. Have we some another conserved quantities? There aren't any from coordinate transformations. May be we have some dynamical ones?

For this purpose let us consider the equation of motion:

$$\frac{d\vec{p}}{dt} = f(r)\frac{\vec{r}}{r}$$

where the radial force is

$$f(r) = -\frac{dV}{dr}$$

Multiplying this equation on momentum vector

$$\dot{\vec{p}} \times \vec{L} = \frac{mf(r)}{r}[\vec{r} \times [\vec{r} \times \dot{\vec{r}}]] = \frac{mf(r)}{r}\{\vec{r}(\vec{r} \cdot \dot{\vec{r}}) - r^2\dot{\vec{r}}\}$$

and performing needed manipulations, we find

$$\dot{\vec{p}} \times \vec{L} = -mf(r)r^2\frac{d}{dt}\left(\frac{\vec{r}}{r}\right)$$

If we take into account the constancy of angular momentum vector, we obtain

$$\frac{d}{dt}[\vec{p}\times\vec{L}] = -mf(r)r^2\frac{d}{dt}\left(\frac{\vec{r}}{r}\right)$$

The further advance is impossible without specifying of $f(r)$. One sees that the integration of this equation becomes possible only if $r^2 f(r) = const.$, or if $f(r) \approx r^{-2}$. But it is the Newton's or Coulomb force

$$f(r) = -\frac{a}{r^2}, \qquad a \equiv GmM \quad \text{or} \quad kq_1q_2.$$

In this case the above equation takes the form:

$$\frac{d}{dt}\left\{[\vec{p}\times\vec{L}] - ma\frac{\vec{r}}{r}\right\} = 0$$

or we obtain the conserved quantity

$$\vec{A} = [\vec{p}\times\vec{L}] - ma\hat{r} = const,$$

where $\hat{r} \equiv \frac{\vec{r}}{r}$ is an unit vector along the radius-vector $\vec{r}$.

The vector $\vec{A}$ is well-known Runge-Lenz vector, while the priority belongs to Laplace, therefore we call it as Laplace-Runge-Lenz (LRL) vector.

Let consider some of its applications:

*(i)* **Orbit Equation**

It is clear from the definition that this vector lies in the orbit plane and is perpendicular to angular momentum vector: $(\vec{L}\cdot\vec{A}) = (\vec{A}\cdot\vec{L}) = 0$.

If we introduce an $\theta$ angle between this vector and $\vec{r}$, then their scalar product gives

$$rA\cos\theta = (\vec{r}\cdot[\vec{p}\times\vec{L}]) - mar$$

The cyclic permutation into the parenthesis leads to

$$(\vec{r}\cdot[\vec{p}\times\vec{L}]) = (\vec{L}\cdot[\vec{r}\times\vec{p}]) = (\vec{L}\cdot\vec{L}) = L^2$$

Therefore $rA\cos\theta = L^2 - mar$ and we obtain the orbit equation

$$r = \frac{L^2}{ma}\left(1 + \frac{A}{ma}\cos\theta\right)^{-1}$$

Thus the LRL vector gives the possibility to find the orbit equation algebraically without solving to equation of motion.

If we compare this to the conventional form of orbit equation, namely

$$r = P(1 + \varepsilon\cos\theta)^{-1},$$

where P is the so-called orbit parameter, $P = \frac{L^2}{ma}$, we conclude, that

$$A = ma\varepsilon,$$

Here $\varepsilon$ is the orbit eccentricity. $\varepsilon = \frac{A}{ma}$. Reduction to the usual expression is achieved by employing the square of LRL vector:

$$\vec{A}^2 = [\vec{p} \times \vec{L}]^2 + m^2 a^2 - 2ma[\vec{p} \times \vec{L}] \cdot \vec{r} = p^2 L^2 - (\vec{p} \cdot \vec{L})^2 + m^2 a^2 - 2mL^2 \frac{a}{r} =$$

$$= 2mL^2 \left( \frac{p^2}{2m} - \frac{a}{r} \right) + m^2 a^2$$

and because the full energy is $E = \frac{p^2}{2m} - \frac{a}{r}$, we have

$$A^2 = 2mEL^2 + m^2 a^2$$

or

$$\frac{A}{ma} = \sqrt{1 + \frac{2EL^2}{ma^2}},$$

which coincides exactly with the traditional expression of $\varepsilon$.

So, the additional integral of motion gives us the orbit equation.

### (ii) Algebraic aspects of the Kepler Problem

It is interesting to find a symmetry for LRL vector, if any. Let us begin with the angular momentum, $\vec{L}$, which generates 3-space rotations according to the Poisson brackets

$$\{L_i, L_j\} = \varepsilon_{ijk} L_k$$

and, in addition this vector has a vanishing Poisson bracket with the Hamiltonian for any central potential.

It is easy to show that

$$\{L_i, A_j\} = \varepsilon_{ijk} A_k,$$

which expresses the evident fact, that $\vec{A}$ is a vector. The following step is derivable easily

$$\{A_i, A_j\} = -2m|H|\varepsilon_{ijk} L_k$$

Let us chose another normalization for this vector,

$$A_i \to D_i = \frac{1}{\sqrt{-2m|H|}} A_i$$

It is clear that for negative total energies (finite motion) $D_i$ is a real vector and

$$\{D_i, D_j\} = \varepsilon_{ijk} L_k$$

All the three brackets together form the closed algebraic system, which looks better in terms of following linear combinations

$$J_i = \frac{L_i + D_i}{2}, \qquad K_i = \frac{L_i - D_i}{2}.$$

Then the above Poisson brackets transform to

$$\{J_i, J_j\} = \varepsilon_{ijk} J_k, \qquad \{K_i, K_j\} = \varepsilon_{ijk} K_k$$

$$\{J_i, K_j\} = 0$$

These relations express two sets of conserved quantities explicitly, each of them describe some 3-dimensional rotations (the first row), which are mutually independent (the second row).

Therefore we are dealing with the direct product of two O(3), which is equivalent to the rotations in higher, 4-dimensional Euclidean space $O(4) \sim O(3) \times O(3)$.

## 3. Dynamical Symmetry in Quantum Mechanics

It is natural that the dynamical symmetries appear also in quantum mechanics. First, who played attention to this fact, was W. Pauli [5]. He defined the quantum analog of LRL vector as

$$\vec{A} = \frac{1}{2}\left([\vec{p} \times \vec{L}] - [\vec{L} \times \vec{p}]\right) - ma\hat{\vec{r}}$$

where the Weyl ordering is used in transition to quantum mechanics

$$[\vec{p} \times \vec{L}] \rightarrow \frac{1}{2}\left(\vec{p} \times \vec{L} - \vec{L} \times \vec{p}\right)$$

There follows Hermitian operator after such transition.

It is known the commutation relations for orbital momentum operators

$$[L_i, L_j] = i\varepsilon_{ijk} L_k$$

There appear the following relations, when the total energy is negative

$$[L_i, A_j] = i\varepsilon_{ijk} A_k$$
$$[A_i, A_j] = -2m|H| i\varepsilon_{ijk} L_k$$

Therefore we are faced with the extended algebra for operators $L_i$ and $D_i \equiv \left(\sqrt{-2m|H|}\right)^{-1} A_i$, the structure of which is the same as in classical mechanics, but now they are written in the form of commutators. Let define

$$\vec{J}(1) = \frac{1}{2}\left(\vec{L} + \vec{A}\right), \qquad \vec{J}(2) = \frac{1}{2}\left(\vec{L} - \vec{A}\right)$$

Then we have

$$[J_i(a), J_j(a)] = i\varepsilon_{ijk} J_k(a), \qquad (a = 1,2)$$
$$[J_i(1), J_j(2)] = 0$$

So, the algebra is shifted into two SU(2) components, and eigenvalues are

$$\vec{J}^2(1) = j_1(j_1 + 1), \qquad \vec{J}^2(2) = j_2(j_2 + 1) \qquad j_i = 0, 1/2, 1, \ldots$$

But because of orthogonally $(\vec{L} \cdot \vec{A}) = (\vec{A} \cdot \vec{L}) = 0$, their eigenvalues (Casimir operators) are the same

$$\vec{J}^2(1) = \vec{J}^2(2) = \frac{1}{4}\left(\vec{L}^2 + \vec{A}^2\right) \rightarrow j(j+1), \qquad j \equiv j_1 \equiv j_2$$

In the case of negative energies we have (in suitable units)

$$H = -\frac{1}{2\left(\vec{L}^2 + \vec{D}^2 + 1\right)^2}$$

Therefore we find for energy spectrum

$$E = -\frac{1}{2}\frac{1}{(2j+1)^2}$$

which is nothing but the Ballmer's known formula for the hydrogen atom spectrum, if the main quantum number is identified as $n = 2j+1 = 0,1,2,...$

The same result follows from the Schrodinger equation. The characteristic feature is the degeneracy with respect to $l$. This parameter enters the Schrodinger equation explicitly but energy does not depend on it. This phenomenon is named as an "accidental" degeneracy. The reason of that degeneracy is the peculiarity of the Coulomb potential. V.Fock[6] was the first, who opened this secret. He took the attention to the additional O(4) symmetry in the momentum representation, which is connected to the LRL vector.

As we have seen above, the symmetry considerations correctly describe the hydrogen atom. The spectrum is derived algebraically without solving of the Schrodinger equation. We see, that while we still have an algebra in quantum mechanics, nevertheless is not known again what transforms by LRL vector.

## 4. Relativistic Quantum Mechanics ( Dirac Equation)

In 1916 A.Sommerfeld [7] derived the formula for the energy spectrum of hydrogen atom in relativistic mechanics using his quasiclassical quantization method. This formula looks like:

$$E = m\left\{1 + \frac{(Z\alpha)^2}{\left(n - |\kappa| + \sqrt{\kappa^2 - (Z\alpha)^2}\right)}\right\}^{-1/2}$$

Here $|\kappa| = j+1/2$. After several years (10-12) Dirac [8] introduced the correct relativistic equation, in which the spin of electron arises automatically. Dirac solved this equation himself for hydrogen atom with the Coulomb potential. His result coincides to the Sommerfeld's one. This paradox has been solved only after 30-40 years later [9]. The fact is that the hydrogen atom spectrum is given by Sommerfeld formula. This formula removes all the degeneracies , which took place in NR quantum mechanics, but one degeneracy still remains – the spectrum depends only on the eigenvalues of total momentum $\vec{J} = \vec{L} + \vec{S}$ (or on its eigenvalues $j$). Therefore there remains a twofold degeneracy for $\kappa = \pm(j+1/2)$, according of which levels $E(S_{1/2})$ and $E(P_{1/2})$ must be degenerate, but experimentally small shift was founded by Lamb [10]. This level shift is named as the Lamb shift. It was explained in QED [11]. It seems that one of the motivation of creation of the quantum electrodynamics was the aspiration for the explanation of the Lamb shift [12].

The natural question appears – Is there some symmetry in the Dirac equation behind the prohibition of the Lamb shift?

Below we shell see that the answer to this question is affirmative.

Let us consider the general Dirac Hamiltonian

$$H = \vec{\alpha} \cdot \vec{p} + \beta m + V(r)$$

where $V(r)$ is an arbitrary central potential, which is the 4th component of the Lorentz vector, in accordance with the minimal gauge switching. This Hamiltonian commutes with the total momentum operator $\vec{J} = \vec{L} + \frac{1}{2}\vec{\Sigma}$, where $\vec{\Sigma} = diag(\vec{\sigma}, \vec{\sigma})$ is the spin matrix of fermions. It is easy to confirm, that the following operator $K = \beta(\vec{\Sigma} \cdot \vec{L} + 1)$ commutes with H. It is named as Dirac operator.

The eigenvalue of this operator is exactly $\kappa$, mentioned above and the degeneracy with respect of $\kappa$ remains in the solution of Coulomb problem.

Let us remark that:

$\kappa = j + 1/2$, when $j = l + 1/2$, leading to levels ($S_{1/2}$, $P_{3/2}$, etc.) and

$\kappa = -(j + 1/2)$, when $j = l - 1/2$ $\Rightarrow$ ($P_{1/2}$, $D_{3/2}$, etc.).

Therefore the forbidden of the Lamb shift results from $\kappa \to -\kappa$ symmetry which at the same time means the reflection of the spin direction with respect to the angular momentum direction. Let us find the operator which reflects this sign. It's evident that such an operator, say $Q_1$, if it exists, must be anticommuting with K,

$$\{Q_1, K\} \equiv Q_1 K + K Q_1 = 0$$

It is clear that the following operator

$$Q_2 = i \frac{Q_1 K}{\sqrt{K^2}}$$

would be anticommuting with K and $Q_1$ as well. Moreover the introduced operators have equal squares

$$\{Q_1, Q_2\} = 0, \qquad Q_1^2 = Q_2^2 \equiv \tilde{H}$$

One can now construct new operators

$$Q_\pm = Q_1 \pm i Q_2$$

They are nilpotent, $Q_\pm^2 = 0$ and $\{Q_+, Q_-\} = 2\tilde{H}$, $[Q_i, \tilde{H}] = 0$.

These algebraic relations define the structure, which is called as Witten's algebra or N=2 superalgebra [13]. (There are anticommutators together with commutators in superalgebras. Such structures in mathematics are known as graded Lie algebras).

What happens if we require invariance of Dirac Hamiltonian with respect to this algebra? Or if we require

$$[Q_i, H] = 0, \qquad i = 1, 2$$

Thus, we are faced to the following problem: **Find (construct) the operator(s), which anticommutes with the Dirac K operator and commutes with the Dirac Hamiltonian, H.**

Let us first construct the anticommuting operator. One of such operator is Dirac's $\gamma^5$ matrix. What else? There is a simple Theorem [14]:

*If $\vec{V}$ is a vector with respect of angular momentum operator $\vec{L}$, i.e.*
$$[L_i, V_j] = i\varepsilon_{ijk} V_k$$
*and simultaneously it is perpendicular to it,* $(\vec{L} \cdot \vec{V}) = (\vec{V} \cdot \vec{L}) = 0$,

*then the following operator* $(\vec{\Sigma} \cdot \vec{V})$, *which is scalar with respect of total momentum* $\vec{J} = \vec{L} + \frac{1}{2}\vec{\Sigma}$, *anticommutes with K:*

$$\{K, (\vec{\Sigma} \cdot \vec{V})\} = 0$$

*In general,*

$$\{K, \hat{O}(\vec{\Sigma} \cdot \vec{V})\} = 0,$$

*where* $\hat{O}$ *is commuting with K.*

Armed by this theorem, one can choose physical vectors at hand, which obey the conditions of this Theorem. They are:

$\vec{V} = \hat{\vec{r}}$ - Unit radius-vector,

and $\vec{V} = \vec{p}$ - linear momentum vector.

There is also the LRL vector. But its inclusion is not needed, because

$$(\vec{\Sigma} \cdot \vec{A}) = \vec{\Sigma} \cdot \hat{\vec{r}} + \frac{i}{ma}\beta K(\vec{\Sigma} \cdot \vec{p}).$$

and the relevant operator is expressible via known operators, owing the relation

$$K(\vec{\Sigma} \cdot \vec{V}) = -i\beta\left(\vec{\Sigma} \cdot \frac{1}{2}\{[\vec{V} \times \vec{L}] - [\vec{L} \times \vec{V}]\}\right)$$

Therefore we construct the most general K-odd operator in the following form

$$Q_1 = x_1(\vec{\Sigma} \cdot \hat{\vec{r}}) + ix_2 K(\vec{\Sigma} \cdot \vec{p}) + ix_3 K\gamma^5 f(r)$$

Now requiring the commutativity with the Hamiltonian,

$$[Q_1, H] = (\vec{\Sigma} \cdot \hat{\vec{r}})\{x_2 V'(r) - x_3 f'(r)\} + 2i\beta K\gamma^5\left\{\frac{x_1}{r} - mf(r)x_3\right\} = 0,$$

it follows the relations:

$$x_2 V'(r) = x_3 f'(r)$$

$$x_3 mf(r) = \frac{x_1}{r}$$

Then we find

$$V(r) = \frac{x_1}{x_2}\frac{1}{mr}$$

So, we can conclude, that the only central potential for which the Dirac Hamiltonian is supersymmetric in the above sense, is the Coulomb one.

### *The Physical Meaning of Constructed Conserved Operator*

To elucidate the physical meaning of derived operator let make use of obtained relations and rewrite it to more transparent form by application of Dirac's algebra:

$$Q_1 = \vec{\Sigma} \cdot \left\{\hat{\vec{r}} - \frac{i}{2ma}\beta([\vec{p} \times \vec{L}] - [\vec{L} \times \vec{p}])\right\} + \frac{i}{mr}K\gamma^5$$

Lippmann and Johnson [15] in 1950 published this form in a brief abstract, where it is said only, that this operator commutes with the Dirac Hamiltonian in Coulomb potential and replaces the LRL vector, known in NR quantum mechanics. But by unknown reason they never published the derivation of this operator (very curious fact in the history of 20$^{th}$ century physics). It was a reason, perhaps, that our article was published right a way in 2005 [14].

For our aim it is useful to perform non-relativistic limits $\beta \to 1$, $\gamma^5 \to 0$. Then it remains

$$Q_1 \to \vec{\Sigma} \cdot \vec{A},$$

where $\vec{A}$ is the LRL vector, i.e. this operator turns into the spin projection of the LRL vector.

The Lamb shift is explained in QED by taking into account the radiative corrections in the photon propagator and photon-electron vertex, which gives the following additional piece in Hamiltonian [11]:

$$\Delta V_{Lamb} \approx \frac{4\alpha^2}{3m^2}\left(\ln\frac{m}{\mu} - \frac{1}{5}\right)\delta^{(3)}(\vec{r}) + \frac{\alpha^2}{2\pi m^2 r^3}\left(\vec{\Sigma} \cdot \vec{L}\right)$$

This expression does not commute with our obtained Johnson-Lippmann (JL) operator. Therefore when only Coulomb potential is considered in the Dirac equation as in an one-electron theory, Lamb shift would be always forbidden.

We see that the hidden symmetry of the Coulomb potential governs the physical phenomena in a sufficiently wide interval – from planetary motion to the fine and hyperfine structure of atomic spectra.

### *Calculation of the Hydrogen Atom Spectrum*

For this aim let us calculate the square of obtained conserved operator by analogy of classical mechanics. This gives [14]

$$Q_1^2 = 1 + \left(\frac{K}{a}\right)^2\left(\frac{H^2}{m^2} - 1\right)$$

All the operators entering here commute with each others. Therefore one can replace them by corresponding eigenvalues and then solve from it for energy. Because of positive definiteness of $Q_1^2$ as the square of Hermitian operator, its minimal quantity is zero. This gives for the ground state energy

$$E_0 = m\left(1 - \frac{(Z\alpha)^2}{\kappa^2}\right)^{1/2}$$

The full spectrum follows from this expression by using the well-known step procedure [13], which reduces in our case to the following substitution

$$\sqrt{\kappa^2 - a^2} \to \sqrt{\kappa^2 - a^2} + n - |\kappa|, \qquad a \equiv Z\alpha.$$

Then the Sommerfeld formula follows.

Thus, in case of Dirac equation the spectrum of the hydrogen atom is obtainable from the symmetry considerations alone.

We see that the supersymmetry requirement appears to be a very strong constraint in the framework of the Dirac Hamiltonian. While the supercharge operator, commuting with the Dirac Hamiltonian in case of pure vector component only is intimately related to the LRL vector, but unlike the latter one relativistic supercharge participates in transformations of spin degrees of freedom. In passing to non-relativistic physics, information concerning to spin–degrees of freedom disappears and hence LRL vector as a generator of algebra does not transform anything and symmetry becomes hidden as a relic of relativistic quantum mechanics.

### *Lorentz-Scalar Potential in the Dirac Equation*

Another important example, which corresponds to non–minimal coupling in the Dirac equation, is a Lorentz–scalar potential. This potential together with the 4$^{th}$ component of Lorentz–vector completes most general central interaction in the Dirac equation.

Let us consider the full Hamiltonian

$$H = \vec{\alpha} \cdot \vec{p} + \beta m + V(r) + \beta S(r)$$

Here $S(r)$ is a Lorentz–scalar, while $V(r)$ is only $O(3)$ scalar, but a 4$^{th}$ component of Lorentz–vector.

Now this Hamiltonian still commutes with Dirac's $K$-operator, but does not commute with the above JL operator even for Coulomb potential.

It is evident that non–relativistic quantum mechanics is indifferent as regards to the Lorentz transformation (variance) properties of potential, therefore it is expected that in case of scalar potential the description of hidden symmetry should also be possible. In other words, the JL operator must be generalized.

For this purpose we make use of our method based on the theorem about $K - odd$ structures. As compared with the previous case, one chooses the generalization, particularly in the part of additional $\hat{O}$ factors.

Therefore we probe the following operator

$$Q_1 \equiv X = x_1(\vec{\Sigma} \cdot \hat{\vec{r}}) + x_1'(\vec{\Sigma} \cdot \hat{\vec{r}})H + ix_2 K(\vec{\Sigma} \cdot \vec{p}) + ix_3 K\gamma_5 f_1(r) + ix_3' K\gamma_5 \beta f_2(r)$$

We have included $\hat{O} = H$ in the first structure and at the same time the matrix $\hat{O} = \beta$ in the third one. Both of them commute with $K$. It is a minimal extension of the previous picture, when only the first orders structures in $\hat{\vec{r}}$ and $\vec{p}$ participate. For turning to the previous case one must take $x_1' = x_3' = 0$ and $S(r) = 0$.

Calculations of relevant commutators give

$$[X,H] = \gamma^5 \beta K \left\{ \frac{2ix_1}{r} - 2ix_3(m+S)f_1(r) + \frac{2ix_1'}{r}V(r) \right\} +$$

$$+ K(\vec{\Sigma} \cdot \hat{r})\{x_2 V'(r) - x_3 f_1'(r)\} +$$

$$+ K\beta(\vec{\Sigma} \cdot \hat{r})\{x_2 S'(r) - x_3' f_2'(r)\} +$$

$$+ \gamma^5 K \left\{ \frac{2ix_1'(m+S)}{r} - 2ix_3'(m+S)f_2(r) \right\} +$$

$$+ \beta K \left\{ \frac{2ix_1'}{r} - 2ix_3' f_2(r) \right\} (\vec{\Sigma} \cdot \vec{p})$$

Equating this expression to zero, we derive matrix equation, then after passing to $2 \times 2$ representation we must equate to zero separately the coefficients standing in fronts of diagonal and antidiagonal elements.

In this way it follows equations:

1) from diagonal structures ($K(\vec{\Sigma} \cdot \hat{r})$, $K\beta(\vec{\Sigma} \cdot \hat{r})$, $\beta K(\vec{\Sigma} \cdot \vec{p})$):

$$x_2 V'(r) - x_3 f_1'(r) = 0$$

$$\frac{x_1}{r} - x_2(m+S)V(r) + \frac{x_1'}{r}V(r) = 0$$

$$x_2 S'(r) - x_3' f_2'(r) = 0$$

$$\frac{x_1'}{r} - x_3' f_2(r) = 0$$

2) from antidiagonal structures $(\gamma^5 K, \gamma^5 \beta K)$:

$$\frac{x_1}{r} - x_3(m+S)f_1(r) + \frac{x_1'}{r}V(r) = 0$$

$$\frac{x_1'}{r}(m+S) - x_3'(m+S)f_2(r) = 0$$

Integrating the first and third equations of 1) for vanishing boundary conditions at infinity, we obtain

$$f_1(r) = \frac{x_2}{x_3}V(r), \qquad f_2(r) = \frac{x_2}{x_3'}S(r)$$

and taking into account the last equations from 1) and 2), we have

$$f_2(r) = \frac{x_1'}{x_3' r}$$

Therefore, according to previous relation for $S(r)$, we obtain finally

$$S(r) = \frac{x_1'}{x_2 r}$$

So, the scalar potential must be Coulomb.

Inserting now $f_1$ into the first equation of 2) and solving for $V(r)$, one derives

$$V(r) = \frac{x_1}{r} \frac{1}{x_2(m+S) - \frac{x_1'}{r}},$$

At last, using here derived expression for $S(r)$, we find

$$V(r) = \frac{x_1}{x_2 mr}$$

Therefore we make sure that the $N = 2$ supersymmetry in the above described content is symmetry of the Dirac Hamiltonian only for Coulomb potential (for any general combination of Lorentz–scalar and 4$^{th}$ component of a Lorentz–vector).

Now if we take into account above obtained relations and use them into the general expression for $X$, one can reduce it to more compact form

$$X = (\vec{\Sigma} \cdot \hat{\vec{r}})(ma_V + Ha_S) - iK\gamma_5(H - \beta m)$$

where the following notations are used:

$$a_V = -\frac{x_1}{x_2 m}, \qquad a_S = -\frac{x_1'}{x_2}$$

Here $a_i - s$ are the constants of corresponding Coulomb potentials

$$V(r) = -\frac{a_V}{r}, \qquad S(r) = -\frac{a_S}{r},$$

It is remarkable that the above result is derived by general 3–dimensional approach without referring to radial equation, as early. Therefore it is more systematic and transparent.

One comment is necessary to make here:

In this combined case the physical interpretation of conserved operator in terms of LRL vector in NR limit is not so evident. It seems that for this non-minimal generalization the symmetry related to the LRL vector walks a way to the back in favor of supersymmetry.

## Algebraic Derivation of the Spectrum of the Dirac Hamiltonian for an Arbitrary Combination of the Lorentz-Scalar and Lorentz-Vector Coulomb Potential

We have demonstrated above the efficiency of algebraic methods. We elucidated that the Witten's $N = 2$ superalgebra rises immediately as soon as an operator anticommuting with $K$ is constructed. In our case it is only sufficient to introduce the supersymmetric generators as follows

$$Q_1 = X, \qquad Q_2 = i\frac{XK}{|K|}$$

Then the anticommutativity $\{X, K\} = 0$, yields

$$\{Q_1, Q_2\} = 0, \qquad Q_1^2 = Q_2^2 \equiv \tilde{H}$$

Therefore, we are faced again with the Witten's Hamiltonian, $\tilde{H}$.

Now we want to obtain spectrum of the Dirac Hamiltonian pure algebraically, without any referring on equations of motion. Our method is based on Witten's superalgebra, established above.

To explore this algebra, one defines a SUSY ground state $|0\rangle$:

$$\tilde{H}|0\rangle = X^2|0\rangle = 0 \quad \to X|0\rangle = 0$$

Because $X^2$ is a square of Hermitian operator, it has a positive definite spectrum and one is competent to take zero this operator itself in ground state. By this requirement we'll obtain Hamiltonian in this ground state and, correspondingly, ground state energy. After that by well known ladder procedure one can construct the energies of all excited levels.

Let equate $X = 0$ and solve $H$ from Eq. (VI.27)):

$$H = m[(\vec{\alpha}\cdot\hat{\vec{r}})a_S + iK]^{-1}[iK\beta - a_V(\vec{\alpha}\cdot\hat{\vec{r}})] = \frac{m}{\kappa^2 + a_S^2} N$$

where

$$N \equiv [(\vec{\alpha}\cdot\hat{\vec{r}})a_S - iK][iK\beta - a_V(\vec{\alpha}\cdot\hat{\vec{r}})] =$$
$$= -a_S a_V + K[K\beta + ia_V(\vec{\alpha}\cdot\hat{\vec{r}})] - ia_S K\beta(\vec{\alpha}\cdot\hat{\vec{r}})$$

Now we try to diagonalize this operator using Foldy-Wouthuysen [17] like transformation. Because the second and third terms do not commute with each others we need several (at least two) such transformations.

We choose the first transformation in the following manner

$$\exp(iS_1) = \exp\left(-\frac{1}{2}\beta(\vec{\alpha}\cdot\hat{\vec{r}})w_1\right)$$

It is evident that

$$\exp(iS_1)(\vec{\alpha}\cdot\hat{\vec{r}})\exp(-iS_1) = \exp(2iS_1)(\vec{\alpha}\cdot\hat{\vec{r}})$$
$$\exp(iS_1)\beta\exp(-iS_1) = \exp(2iS_1)\beta$$

Moreover

$$\exp(iS_1)K\exp(-iS_1) = K, \qquad \exp(iS_1)\beta K\exp(-iS_1) = \exp(2iS_1)\beta K$$

and $\qquad \exp(iS_1)\beta(\vec{\alpha}\cdot\hat{\vec{r}})\exp(-iS_1) = \beta(\vec{\alpha}\cdot\hat{\vec{r}})$

Therefore the first transformation acts as

$$N' \equiv \exp(iS_1)N\exp(-iS_1) = -a_S a_V + K\exp(2iS_1)[K\beta + ia_V(\vec{\alpha}\cdot\hat{\vec{r}})] - ia_S K\beta(\vec{\alpha}\cdot\hat{\vec{r}})$$

But

$$\exp(2iS_1) = chw_1 + i\beta(\vec{\alpha}\cdot\hat{\vec{r}})shw_1$$

Make use of this relation, we have

$$\exp(2iS_1)[K\beta + ia_V(\vec{\alpha}\cdot\hat{\vec{r}})] =$$
$$= \beta[Kchw_1 + a_V shw_1] + K(\vec{\alpha}\cdot\hat{\vec{r}})[ia_V chw_1 + iKshw_1]$$

Now in order to get rid of non-diagonal $(\vec{\alpha}\cdot\hat{\vec{r}})$ terms, we must choose

$$thw_1 = -\frac{a_V}{K}$$

Using simple trigonometric relations we arrive at

$$\exp(2iS_1)\left[K\beta + ia_V(\vec{\alpha}\cdot\hat{\vec{r}})\right] = K^{-1}\beta\sqrt{\kappa^2 - a_V^2}$$

Let us perform the second F.-W. transformation

$$N'' = \exp(iS_2)N'\exp(-iS_2), \qquad \text{where} \qquad S_2 = -\frac{1}{2}(\vec{\alpha}\cdot\hat{\vec{r}})w_2$$

Now

$$\exp(iS_2)K\beta\exp(-iS_2) = \exp(2iS_2)K\beta$$
$$\exp(iS_2)K\beta(\vec{\alpha}\cdot\hat{\vec{r}})\exp(-iS_2) = \exp(2iS_2)K\beta(\vec{\alpha}\cdot\hat{\vec{r}})$$
$$\exp(2iS_2) = \cos w_2 - i(\vec{\alpha}\cdot\hat{\vec{r}})\sin w_2$$

Therefore

$$N'' = -a_S a_V + K\sqrt{\kappa^2 - a_V^2}\exp(2iS_2)\beta - ia_S\exp(2iS_2)K\beta(\vec{\alpha}\cdot\hat{\vec{r}}) =$$
$$= -a_S a_V + K\sqrt{\kappa^2 - a_V^2}\beta\cos w_2 + iK\sqrt{\kappa^2 - a_V^2}\beta(\vec{\alpha}\cdot\hat{\vec{r}})\sin w_2 -$$
$$- ia_S K\beta(\vec{\alpha}\cdot\hat{\vec{r}})\cos w_2 + a_S K\beta\sin w_2$$

Requiring absence of $(\vec{\alpha}\cdot\hat{\vec{r}})$ terms we have

$$tgw_2 = \frac{a_S}{\sqrt{\kappa^2 - a_V^2}}$$

Therefore

$$N'' = -a_S a_V + K\beta\sqrt{\kappa^2 - a_V^2 + a_S^2}$$

and finally

$$H = \frac{m}{\kappa^2 + a_S^2}\left\{-a_S a_V + K\sqrt{\kappa^2 - a_V^2 + a_S^2}\,\beta\right\}$$

For eigenvalues in ground state we have

$$E_0 = \frac{m}{\kappa^2 + a_S^2}\left\{-a_S a_V \pm \kappa\sqrt{\kappa^2 - a_V^2 + a_S^2}\right\}$$

Now let us remember the result obtained by explicit solution of the Dirac equation for this case [17]

$$E = m\left\{\frac{-a_S a_V}{a_V^2 + (n-|k|+\gamma)^2} \pm \sqrt{\left(\frac{a_S a_V}{a_V^2 + (n-|\kappa|+\gamma)^2}\right)^2 + \frac{(n-|\kappa|+\gamma)^2 - a_S^2}{a_V^2 + (n-|\kappa|+\gamma)^2}}\right\}$$

where

$$\gamma^2 = \kappa^2 - a_V^2 + a_S^2$$

In the ground state $n = 1$, $j = 1/2$ $\rightarrow$ $|\kappa| = j + 1/2 = 1$, there remains

$$E_0 = m\left\{\frac{-a_S a_V}{a_V^2 + \gamma^2} \pm \sqrt{\left(\frac{a_S a_V}{a_V^2 + \gamma^2}\right)^2 + \frac{\gamma^2 - a_S^2}{a_V^2 + \gamma^2}}\right\}$$

This relation after obvious manipulations reduces to our above derived expression.

Therefore by only algebraic methods we have obtained the correct expression for ground state energy.

For obtaining of total spectrum it is sufficient now to use the Witten's algebra. Following to the ordinary step procedure [13], this consists in change (for our case):

$$\gamma \to \gamma + n - |\kappa|$$

Making use of this, it follows the correct expression for total energy spectrum.

Thus, the Dirac equation for the generalized Coulomb potential – an arbitrary combination of Scalar and Lorentz 4-vector – is solved algebraically due to underlying supersymmetry, i.e. this problem is totally integrable [18].

## 5. Generalization to Arbitrary Dimensions

The Dirac equation is analytically solvable in arbitrary (D+1) - dimensions, if the potential is [19]

$$V(r) = -\frac{a}{r},$$

where now $r$ is a radius of the D-dimensional sphere. While studying the Dirac equation in connection with space extension the continuation of Dirac's matrices is necessary as well. It does not meet any difficulties except $\gamma^5$ matrix, which must anticommute with all other gamma matrices in (D+1)-space. As is well known the analogue of this matrix in extended spaces are expressed differently in terms of another gamma-matrices in case of odd- and even-dimensions. But it is essential for our aims only that the relevant matrix, which we denote by $\gamma^{D+2}$, exists and satisfies all relations alike to 4-dimensional space. Namely

$$\left(\gamma^{D+2}\right)^+ = \gamma^{D+2}, \quad \left(\gamma^{D+2}\right)^2 = 1, \quad \{\gamma^{D+2}, \gamma^\mu\} = 0, \quad (\mu = 0,1,2,...,D+1).$$

In these notations the Dirac operator has the form [20]

$$K = \gamma^0 \left\{ \frac{i}{2} \sum_{a \neq b} \gamma^a \gamma^b L_{ab} + \frac{1}{2}(D-1) \right\} =$$

$$= \gamma^0 \left\{ \vec{J}^2 - \vec{L}^2 - \vec{S}^2 + \frac{1}{2}(D-1) \right\}$$

It commutes with the Dirac Hamiltonian in $-a/r$ field

$$H = \gamma^0 m + \gamma^0 \gamma^i p^i - \frac{Z\alpha}{r}, \quad i = 1,2,3,....,D$$

and anticommutes with the generalized JL operator

$$A \equiv \gamma^{D+2} \gamma^0 \gamma^i \frac{x_i}{r} - \frac{i}{ma} K \gamma^{D+2} (H - \gamma^0 m)$$

It is evident that because of similarity of gamma-matrix algebra (we used only algebra in the previous considerations) one can see that for *arbitrary* potential the problem is reduced to the simple repetition of above derived relationships. Therefore we obtain now, that the requirement of the same N=2 supersymmetry will give the following potential

$$V \approx \frac{1}{r}$$

among all central potentials. But it is "Coulomb" potential in multidimensional spaces. Here inverted commas are used because it does not coincide to the corresponding potential, obtained from Gauss law.

This result says that the supersymmetry distinguishes the Coulomb problems in dimensions, different from 3. The equation of motion is solved analytically in any dimensions for $1/r$ potential and the closed orbits are derived in this case in classical mechanics. Still P.Erenfest [21] had shown in 1920 that the potential $r^{2-D}$, obtained from Gauss law, does not give the closed orbits and atom become unstable.

Finally, we conclude, that as the degeneracy was accidental, so is the traditional view about the nature of the Coulomb potential, as it would be connected to the LRL kind of symmetry.

It is manifested in relativistic quantum mechanics, because of peculiar suprsymmetry, which stays long away from the usual ideas about dynamical symmetry of Coulomb potential.

While this result in arbitrary dimensional spaces may be connected to the periodicity of classical orbits, it is cleared up that it, (1/r), governs the wide class of physical phenomena in the large and small distances of the Universe. It seems that this potential could have more deep geometrical background, because its superalgebra now becomes geometric as it performs the reflection of spin.

ACKNOWLEDGEMENTS: One of us (A.K.) Thanks the participants of the Seminar at NYU Drs. G.R.Farrar, G.Gabadadze, D.Zwanziger, V.P.Akulov and others for valuable discussions, as well as Drs. J.Chkareuli, G.Devidze A.N.Kvinikhidze, T.P.Nadareishvili, for many critical comments. This work has been supported by the Georgian National Science Foundation Grant # **GNSF/ST07/4-196.**